\documentstyle[prl,aps]{revtex}
\begin{document}
\draft

\twocolumn[\hsize\textwidth\columnwidth\hsize\csname @twocolumnfalse\endcsname

\title{Off-Diagonal Long-Range Order in Bose Liquids: Irrotational Flow and
Quantization of Circulation}
\author{Gang Su$^{1,2,\ast}$ and Masuo Suzuki$^{2,\dag}$}
\address{$^1$Department of Physics, Graduate School at Beijing,
University of Science and Technology of China, Chinese Academy of Sciences,
P.O. Box 3908, Beijing 100039, China\\
$^2$Department of Applied Physics, Faculty of Science, Science University of Tokyo, 
1-3, Kagurazaki, Shinjuku-ku, Tokyo 162, Japan}
\maketitle

\begin{abstract}
On the basis of gauge invariance, it is proven in an elementary and
straightforward manner, but without invoking any {\it ad hoc} assumption, 
that the existence of off-diagonal long-range order
in one-particle reduced density matrix in Bose liquids implies both the
irrotational flow in a simply connected region and the quantization of
circulation in a multiply connected region, the two fundamental properties
of a Bose superfluid. The origin for both is the phase coherence of
condensate wave-functions. Some relevant issues are also addressed.
\end{abstract}
]

The discovery of superfluidity in liquid $^{4}$He leads to London's proposal
that the transition from He I phase to He II phase is an analog of
Bose-Einstein condensation (BEC), namely, the superfluidity in He II is
characterized by the macroscopic occupation of a single quantum state\cite
{london}, despite that the liquid He II is actually a strongly interacting
Bose system. On the basis of London's idea, Tisza\cite{tisza}, while Landau
independently\cite{landau}, developed phenomenologically a two-fluid model,
in which the superfluid He II is regarded as being composed of superfluid
and normal fluid components. The superfluid component is the single
condensed quantum state, carrying no entropy, and thereby resulting in the
assumption that the superfluid flow is irrotational. Meanwhile, Onsager\cite
{onsager} suggested that quantized vortex lines, or the quantization of
vorticity according to Feynman independently\cite{feynman}, must exist in He
II, which was experimentally verified a few years later\cite{vinen}. Since
then, the fact that the irrotational flow and the quantization of
circulation (vortices) are two fundamental characters of superfluid He II is
well established. Recently, the existence of vortices in a coherent
collection of ultracold atoms has also been proposed and experimentally
observed by spinning up the condensate\cite{willi}. On the other hand,
Penrose and Onsager\cite{penrose} generalized the mathematical description
of BEC to interacting Bose systems, and manifested that BEC appears as the
off-diagonal long-range order (ODLRO) in one-particle reduced density matrix
exists, and liquid helium II was shown to possess such an order. Later, the
concept was successfully extended to superconductivity in interacting
fermion systems by Yang\cite{yang}. It is now well accepted that both
superconductivity and superfluidity are quantum phases characterized by the
existence of ODLRO. Nevertheless, what is the relationship between the
fundamental properties of the condensates and the existence of ODLRO? The
answer to the question for superconductors was presented a few years ago\cite
{sewell,nieh}, i.e., the hypothesis of ODLRO in the two-particle reduced
density matrix implies both Meissner effect and flux quantization, while the
proof for superfluids is still absent.

In this paper, by noting the analog between superconductors and superfluids,
we shall extend the proofs for superconductors presented in Refs. \cite
{sewell,nieh} to interacting Bose systems, and substantiate that the
existence of ODLRO in one-particle reduced density matrix in a Bose liquid
implies both irrotational flow and quantization of circulation. The origin
for the two fundamental properties of a Bose condensate comes from the phase
coherence of condensate wave-functions. Our proof does not invoke any {\it %
ad hoc} assumption.

Let us consider a bucket of Bose liquid composed of $N$ homogeneous
interacting spinless particles, rotated with a constant angular velocity $%
{\bf \Omega }$. The Hamiltonian of the system is 
\begin{equation}
H=\sum_{j}\frac{1}{2m}[{\bf p}_{j}+m{\bf v}_{s}({\bf r}_{j})]^{2}+\frac{1}{2}%
\sum_{j\neq l}V(r_{jl}),  \label{hamil}
\end{equation}
where ${\bf p}_{j}=-i\hbar {\bf \nabla }_{j}$ is the momentum of $j$th
particle, $V(r_{jl})$ denotes the interactions between particles, and ${\bf v%
}_{s}({\bf r}_{j})$ represents the drift velocity, given by 
\begin{equation}
{\bf v}_{s}({\bf r})={\bf \Omega \times r}+{\bf \nabla }\theta ({\bf r}),
\label{veloc}
\end{equation}
in which the first term comes from the Coriolis force, and the second term,
to be identified below, originates from the gauge invariance. Eq. (\ref
{veloc}) comes out from the basic relation ${\bf \Omega =}\frac{1}{2}{\bf %
\nabla }\times {\bf v}_{s}$, as is in a classical case. The Schr\"{o}dinger
equation reads $H\psi _{n}({\bf r}_{1},\cdots ,{\bf r}_{N})=E_{n}\psi _{n}(%
{\bf r}_{1},\cdots ,{\bf r}_{N})$, where the orthonormal eigenfunctions $%
\psi _{n}$'s are single-valued and symmetric. In analogy to the spirit of
Refs. \cite{sewell,nieh}, we first make a gauge transformation. Under a
small spatial displacement ${\bf r\rightarrow r}-{\bf l}$, we have ${\bf v}%
_{s}({\bf r}){\bf \rightarrow v}_{s}({\bf r}-{\bf l})={\bf v}_{s}({\bf r})+%
{\bf \nabla \lbrack l}\cdot ({\bf \Omega \times r)+}\theta ({\bf r}-{\bf l}%
)-\theta ({\bf r})]$, and $V(r_{jl})\rightarrow V(r_{jl})$. Then, the
Schr\"{o}dinger equation becomes 
\begin{equation}
H{\tilde{\psi}}_{n}({\bf r}_{1},\cdots ,{\bf r}_{N})=E_{n}{\tilde{\psi}}_{n}(%
{\bf r}_{1},\cdots ,{\bf r}_{N})  \label{schrodinger}
\end{equation}
where 
\begin{equation}
{\tilde{\psi}}_{n}({\bf r}_{1},\cdots ,{\bf r}_{N})=e^{i\frac{m}{\hbar }%
\sum_{j}\xi ({\bf r}_{j},{\bf l})}\psi _{n}({\bf r}_{1}-{\bf l},\cdots ,{\bf %
r}_{N}-{\bf l})  \label{eigenfunc}
\end{equation}
is also the eigenfunctions of $H$, and the phase factor $\xi ({\bf r},{\bf l}%
)$ is defined by 
\begin{equation}
\xi ({\bf r},{\bf l})={\bf l}\cdot ({\bf \Omega \times r)+}\theta ({\bf r}-%
{\bf l})-\theta ({\bf r}).  \label{phase}
\end{equation}
Therefore, we have another set of orthonormal eigenfunctions, ${\tilde{\psi}}%
_{n}$'s, for the system.

To gain insight into the underlying physics behind ODLRO in Bose liquids, we
ought first to look at the intrinsic properties of the one-particle reduced
density matrix $\rho _{1}({\bf r},{\bf r}^{\prime })$. In accordance with
the definition, $\rho _{1}({\bf r},{\bf r}^{\prime })$ can be written down
as 
\begin{eqnarray}
&&\rho _{1}({\bf r},{\bf r}^{\prime })=\int \cdots \int \frac{d{\bf r}%
_{2}\cdots d{\bf r}_{N}}{(N-1)!}\frac{1}{Z}\times  \nonumber \\
&&\sum_{n}e^{-E_{n}/k_{B}T}\psi _{n}({\bf r},{\bf r}_{2},\cdots ,{\bf r}%
_{N})\psi _{n}^{\ast }({\bf r}^{\prime },{\bf r}_{2},\cdots ,{\bf r}_{N}),
\label{rho-1}
\end{eqnarray}
where $Z=\sum_{n}e^{-E_{n}/k_{B}T}$, is the partition function. $\rho _{1}(%
{\bf r},{\bf r}^{\prime })$ is normalized as $Tr\rho _{1}=\int d{\bf r}\rho
_{1}({\bf r},{\bf r})=N$, and is single-valued. Now, we utilize the energy
eigenfunctions given by Eq. (\ref{eigenfunc}) to represent $\rho _{1}({\bf r}%
,{\bf r}^{\prime })$ anew, and get the relation 
\begin{equation}
\rho _{1}({\bf r},{\bf r}^{\prime })=e^{i\frac{m}{\hbar }[\xi ({\bf r},{\bf l%
})-\xi ({\bf r}^{\prime },{\bf l})]}\rho _{1}({\bf r-l},{\bf r}^{\prime }-%
{\bf l}).  \label{rho-2}
\end{equation}
Equation\ (\ref{rho-2}) illustrates that under a spatial displacement the
one-particle reduced density matrix gains a position and displacement
dependent phase factor. It is an intrinsic property of $\rho _{1}({\bf r},%
{\bf r}^{\prime })$. On the other hand, we can express $\rho _{1}({\bf r},%
{\bf r}^{\prime })$ in the spectral form as 
\begin{equation}
\rho _{1}({\bf r},{\bf r}^{\prime })=\sum_{\nu }\lambda _{\nu }\Phi _{\nu }(%
{\bf r})\Phi _{\nu }^{\ast }({\bf r}^{\prime }),  \label{rho-3}
\end{equation}
where $\Phi _{\nu }$ is the eigenfunctions of $\rho _{1}$, with eigenvalues $%
\lambda _{\nu }$. According to Penrose and Onsager\cite{penrose}, the
existence of ODLRO in $\rho _{1}$ implies the spectral resolution such that 
\begin{equation}
\rho _{1}({\bf r},{\bf r}^{\prime })=\lambda _{0}\Phi _{0}({\bf r})\Phi
_{0}^{\ast }({\bf r}^{\prime })+\rho _{1}^{\prime }({\bf r},{\bf r}^{\prime
}),  \label{spectral}
\end{equation}
where $\lambda _{0}$, on the order of $O(N)$, is the largest eigenvalue of $%
\rho _{1}$, $\Phi _{0}({\bf r})$ is the corresponding eigenfunction (the
so-called condensate wave-function), and $\rho _{1}^{\prime }({\bf r},{\bf r}%
^{\prime })$ denotes other terms being on the order of $O(1)$. In the
thermodynamic limit, $\rho _{1}^{\prime }({\bf r},{\bf r}^{\prime
})\rightarrow 0$ as $|{\bf r}-{\bf r}^{\prime }|\rightarrow \infty $.
Comparing Eqs. (\ref{rho-2}) and (\ref{spectral}) we may obtain under the
limit $|{\bf r}-{\bf r}^{\prime }|\rightarrow \infty $ the following
expression 
\begin{equation}
\Phi _{0}({\bf r})\Phi _{0}^{\ast }({\bf r}^{\prime })=e^{i\frac{m}{\hbar }%
[\xi ({\bf r},{\bf l})-\xi ({\bf r}^{\prime },{\bf l})]}\Phi _{0}({\bf r-l}%
)\Phi _{0}^{\ast }({\bf r}^{\prime }-{\bf l}).  \label{phi-phi}
\end{equation}
With a careful analysis, one may find that this equation suggests 
\begin{equation}
\Phi _{0}({\bf r})=e^{i\zeta ({\bf l)}}e^{i\frac{m}{\hbar }\xi ({\bf r},{\bf %
l})}\Phi _{0}({\bf r-l}),  \label{phi}
\end{equation}
where a displacement dependent, but position independent phase factor, $%
\zeta ({\bf l)}$ with $\zeta ({\bf 0)}=0$, may possibly emerge when
extracting the information of individual function from the products of Eq. (%
\ref{phi-phi}). As a result, we observe that the condensate wave-function, $%
\Phi _{0}({\bf r})$, is mutually related by a phase factor at different
positions. It is this property, being of fundamental importance, that
enables us to prove the irrotational flow and the quantization of
circulation in Bose superfluids.

In a simply connected region inside the superfluid, according to the rule
governed by Eq. (\ref{phi}) we perform two opposite displacements, first $%
{\bf a}$ followed by ${\bf b}$, and then ${\bf b}$ followed by ${\bf a}$.
Through comparing the relations satisfied by $\Phi _{0}({\bf r})$ in the two
operations, we discover that the phase factors satisfy the equation: $F({\bf %
r},{\bf a},{\bf b})=F({\bf r},{\bf b},{\bf a})$, with the function $F({\bf r}%
,{\bf a},{\bf b})=\exp \{i\frac{m}{\hbar }[\xi ({\bf r},{\bf a})+\xi ({\bf r}%
-{\bf a},{\bf b})]\}$. By solving this equation, we have $\exp [i\frac{2m}{%
\hbar }{\bf \Omega \cdot (b\times a})]=1$, yielding 
\begin{equation}
{\bf \Omega \cdot (b\times a})=n\frac{h}{2m},\text{ \ \ }(n=%
\mathop{\rm integer}%
).  \label{omega}
\end{equation}
It is seen that the right-hand side of Eq.(\ref{omega}) is discretized,
while the left-hand side can be continuously changed as the two operations
can be arbitrarily choosen. The consistency condition requires 
\begin{equation}
{\bf \Omega =0.}  \label{omega-0}
\end{equation}
This proves the property of the irrotational flow of the Bose condensate.
However, we should stress that in terms of Eq. (\ref{veloc}) the flow
velocity, i.e., the superfluid velocity, should be in principle
nonvanishing, i.e., ${\bf v}_{s}({\bf r})={\bf \nabla }\theta ({\bf r})$,
satisfying $\nabla \times {\bf v}_{s}({\bf r})=0.$

In a multiply connected region inside the superfluid, there should be some
singularities around which the fluid is rotating. These singularities are
nothing but the vortex lines. In this case, we can repeat successively for
infinitesimal displacements along a closed path $C$ which encloses at least
one vortex line, and then pick up products of phase factors. Consequently,
we have from Eq.(\ref{phi}) the relation $\Phi _{0}({\bf r})=e^{i\frac{m}{%
\hbar }\oint_{C}{\bf v}_{s}\cdot d{\bf l}}\Phi _{0}({\bf r})$, where use has
been made of the single-valuedness of $\Phi _{0}({\bf r})$, being a
consequence of the single-valuedness of $\rho _{1}$, and $\exp [i\zeta
(\oint_{C}d{\bf l})]=\exp [i\zeta ({\bf 0)]=}1$. Therefore, we get 
\begin{equation}
\oint_{C}{\bf v}_{s}\cdot d{\bf l}=n\frac{h}{m},\text{ \ \ }(n=%
\mathop{\rm integer}%
).  \label{circulation}
\end{equation}
This is the quantization of circulation, with the quantum of vorticity being 
$h/m$. If there is no any vortices encircled by the path loop $C$ in the
multiply connected system, then the circulation will be identically zero,
leading to again the property of irrotational flow everywhere away from the
vortex cores.

The irrotational flow and the quantization of circulation have been well
understood for superfluid He II based on the work of London, Onsager and
Feynman quite long time before. Here we have offered another route, without 
{\it ad hoc} assumption, to understand the two fundamental characters of a
Bose superfluid, and then established a deep connection between the basic
properties of the condensate and the existence of ODLRO in Bose liquids. It
is manifested that the condensate, as a single quantum state possessing
these two essential features, is indeed characterized by the eigenstate of
the one-particle reduced density matrix with the largest eigenvalue of the
order of $O(N)$, as argued by Penrose and Onsager\cite{penrose}. The
preceding discussion shows that the phase coherence of condensate
wave-functions is the origin for occurence of both properties.

Next, let us clarify the physical meaning of the quantity $\theta ({\bf r}).$
In general, the condensate wave-function is assumed to be a complex number,
and is usually denoted by 
\begin{equation}
\Phi _{0}({\bf r})=\phi _{0}({\bf r})e^{i\alpha ({\bf r})},
\label{phi-alpha}
\end{equation}
where the amplitude $\phi _{0}({\bf r})$ and the phase factor $\alpha ({\bf r%
})$ are real functions. Define the condensate density as $\rho _{0}({\bf r}%
)\equiv |\Phi _{0}({\bf r})|^{2}=[\phi _{0}({\bf r})]^{2}$. Then, the
superfluid current density at position ${\bf r}$ is given by $J_{0}({\bf r}%
)=-\frac{i\hbar }{2m}(\Phi _{0}^{\ast }\nabla \Phi _{0}-\Phi _{0}\nabla \Phi
_{0}^{\ast })=\frac{\hbar }{m}[\phi _{0}({\bf r})]^{2}\nabla \alpha ({\bf r}%
) $. Since the current density and the superfluid velocity is related by $%
J_{0}({\bf r})=\rho _{0}({\bf r}){\bf v}_{s}({\bf r})$, combining the
irrotational property of the condensate we have 
\begin{equation}
\theta ({\bf r})=\frac{\hbar }{m}\alpha ({\bf r}).  \label{theta}
\end{equation}
As a result, we arrive at the conclusion that the superfluid velocity in the
Bose liquids is proportional to the gradient of the phase factor of the
condensate wave-function, namely 
\begin{equation}
{\bf v}_{s}({\bf r})=\frac{\hbar }{m}\nabla \alpha ({\bf r}),
\label{veloc-1}
\end{equation}
a well-known result that was supposed for superfluid He II by London,
Onsager and Feynman. However, our derivation shows that Eq. (\ref{veloc-1})
holds true for any Bose system so long as the system possesses ODLRO in $%
\rho _{1}$. Thus, the phase of the condensate is closely related to the
gauge invariance.

To explore the displacement dependent but position independent phase factor $%
\zeta ({\bf l})$, let us compare Eqs.\ (\ref{phi}) and (\ref{phi-alpha}). By
considering (\ref{phase}) and (\ref{theta}) we can get $\tan [2\alpha ({\bf r%
})]=\tan [2\alpha ({\bf r}-{\bf l})+\zeta ({\bf l})+\frac{m}{\hbar }{\bf l}%
\cdot ({\bf \Omega }\times {\bf r})]$. Solving this equation and noting $%
\zeta (0)=0$, one may obtain 
\[
\zeta ({\bf l})=2[\alpha ({\bf r})-\alpha ({\bf r}-{\bf l})]-\frac{m}{\hbar }%
{\bf l}\cdot ({\bf \Omega }\times {\bf r}). 
\]
This suggests that such a phase factor indeed exists if the condensate
wave-function takes the form of (\ref{phi-alpha}). Consequently, we have
from Eqs.\ (\ref{phi}) and (\ref{phi-alpha}) 
\begin{equation}
\phi _{0}({\bf r})=\phi _{0}({\bf r}-{\bf l}).  \label{phi0}
\end{equation}
It shows that the amplitude of condensate wave-function, under assumption of
the form (\ref{phi-alpha}), is displacement-invariant. This is convincing,
because in superfluid He II, $\phi _{0}({\bf r})$ is supposed as the square
root of $n_{0}$, the macroscopic occupation number of the condensate, which
is certainly translation-invariant. Here we have proved that for any
condensed Bose system the condensate density possesses a translational
invariance.

In the Bose condensate, the continuity equation becomes $\partial \phi
_{0}/\partial t+{\bf v}_{s}\cdot \nabla \phi _{0}+(1/2)\phi _{0}\nabla \cdot 
{\bf v}_{s}=0.$ After integrating it, we have 
\begin{eqnarray*}
\frac{\partial }{\partial t}\int d{\bf r}\phi _{0}({\bf r},t) &=&\frac{1}{2}%
\int d{\bf r}\phi _{0}({\bf r},t)\nabla \cdot {\bf v}_{s}({\bf r}) \\
&&-\int d{\bf r}\nabla \cdot \lbrack {\bf v}_{s}({\bf r})\phi _{0}({\bf r}%
,t)].
\end{eqnarray*}
In view of Gaussian theorem, the second term of the right-hand side of above
equation should be a surface integral, and is vanishing in the present case.
For a steady flow, $\nabla \cdot {\bf v}_{s}({\bf r})=0$, the above equation
tells us that $\int d{\bf r}\phi _{0}({\bf r},t)$ is independent of time. If
we define the spatial Fourier transform of $\phi _{0}({\bf r},t)$ as $\phi
_{0}({\bf k},t)$, we know that $\lim_{{\bf k}\rightarrow 0}\phi _{0}({\bf k}%
,t)$ is in fact independent of $t$. Consequently, the temporal Fourier
transform of $\phi _{0}({\bf k},t)$, defined by $\phi _{0}({\bf k},\omega )$%
, has a property: $\lim_{{\bf k}\rightarrow 0}\phi _{0}({\bf k},\omega )\sim
\delta (\omega )$. This implies that in the long wavelength limit there must
be excitations whose frequency is zero in the Bose condensate. These
massless excitations are nothing but Goldstone modes. As one knows, in He II
the Goldstone bosons are identified as second sound. Therefore, by means of
Goldstone's theorem one can conclude that in the Bose condensate the gauge
symmetry is broken.

It is worth noticing that the irrotational flow and the quantization of
circulation in a Bose superfluid can also be derived starting directly from
Eqs. (\ref{phi-alpha}) and (\ref{veloc-1}), as was done in previous
literature. However, if we take a careful look, we can observe that the
prerequisite of such reasonings is that the condensate wave-function was
supposed to take the form of Eq. (\ref{phi-alpha}). Contrary to the previous
arguments in which Eq. (\ref{phi-alpha}) was only an assumption without
proof, the present derivation does not depend on what form of the condensate
wave-function is taken, and only invokes the general principle of gauge
invariance. In some sense, our result suggests implicitly that the
condensate wave-function has the form of Eq. (\ref{phi-alpha}).

To this end, since superfluids and superconductors share some common
features, it would be interesting to compare the correspondence between
characters and variables of both, as summarized in the following table.

\begin{center}
\begin{tabular}{c|c}
\hline
Superfluids & Superconductors \\ \hline
Coriolis Force & Electromagnetic force \\ 
Angular velocity (${\bf \Omega }$) & Magnetic field (${\bf B}$) \\ 
Flow velocity (${\bf v}_{s}({\bf r})$) & Vector potential (${\bf A}({\bf r})$%
) \\ 
ODLRO in $\rho _{1}$ & ODLRO in $\rho _{2}$ \\ 
Irrotational flow & Meissner effect \\ 
(${\bf \Omega }=\frac{1}{2}\nabla \times {\bf v}_{s}=0$) & (${\bf B}=\nabla
\times {\bf A}=0$) \\ 
Quantization of circulation & Flux quantization \\ 
($\oint {\bf v}_{s}\cdot d{\bf l}=n\frac{h}{m}$) & ($\oint {\bf A}\cdot d%
{\bf l}=n\frac{ch}{2e}$) \\ 
Broken gauge symmetry & Broken U(1) gauge symmetry \\ 
Phase coherence & Phase coherence \\ 
BEC state of bosons & Condensate of Cooper Pairs \\ \hline
\end{tabular}
\end{center}

\noindent As is seen, the similarity between both systems is a consequence
of ODLRO, or equivalently, of the broken gauge symmetry, which can be viewed
as the profound underlying physics behind these two important systems in
condensed matter, as emphasized by many authors (see, e.g. Ref.\cite{bec}
for more references).

We close this paper with a few remarks in order.

(1) The present proof works for both noninteracting and interacting Bose
systems. Thus, for an ideal Bose gas, as there is an ODLRO in $\rho _{1}$,
the BEC state should also possess the property of irrotational flow and
quantization of circulation. An ideal Bose gas is a superfluid, but it does
not exhibit superfluidity, because it is unstable against any motion of the
system. It is also manifested by Landau's criterion for superfluidity, i.e.
the critical velocity is zero.

(2) For fragmented condensates, namely there are more than one eigenvalues
of the one-particle reduced density matrix, $\rho _{1}$, of the order $O(N)$%
, leading to two or more eigenstates of $\rho _{1}$ being macroscopically
occupied\cite{noz}, our proof seems not to be sufficient, though the forms
like Eq. (\ref{phi}) for condensates are still a possible solution. The
fragmentation of condensates usually occurs in multi-component Bose systems,
and cannot happen in single component systems, because it costs a
macroscopic extensive exchange energy. For non-fragmented systems our
argument is closely related to the so-called phase locking, as discussed by
Nozieres\cite{noz}.

(3) For the Bose systems with spin nonzero, the one-particle reduced density
matrix, $\rho _{1}({\bf r},\sigma ;{\bf r}^{\prime },\sigma ^{\prime })$
with $\sigma $ and $\sigma ^{\prime }$ spin indices, has a spectral
resolution such that: $\rho _{1}({\bf r},\sigma ;{\bf r}^{\prime },\sigma
^{\prime })=\lambda _{0}\Phi _{0}({\bf r,}\sigma )\Phi _{0}^{\ast }({\bf r}%
^{\prime },\sigma ^{\prime })+\rho _{1}^{\prime }({\bf r},\sigma ;{\bf r}%
^{\prime },\sigma ^{\prime })$, where $\rho _{1}^{\prime }\rightarrow 0$ for 
$|{\bf r}-{\bf r}^{\prime }|\rightarrow \infty $. Obviously, the condensate
wave-function, $\Phi _{0}({\bf r,}\sigma )$, is a spinor. In the absence of
external magnetic fields, our reasonings remain similar to those for
spinless systems. However, if a magnetic field is applied to the system,
owing to a so-called ``local spin-gauge symmetry'' \cite{ho} there is a
vorticity induced by spatial variations of the magnetic field, and the
relation like Eq. (\ref{veloc}) no longer holds, suggesting that our proof
appears to be invalid, which is a possible consequence of the fact that the
time-reversal symmetry is violated in the presence of an external field.

(4) For trapped alkali atoms like $^{7}$Li, $^{23}$Na, $^{39}$K, $^{87}$Rb
etc, the signature of BEC is successfully observed at extremely low
temperatures (see, e.g. Ref.\cite{dalvo} for a review). These trapped Bose
gases have finite sizes and are inhomogeneous, which contain the number of
atoms typically from a few thousands to several millions. Therefore, the
thermodynamic limit is not exactly reached for trapped gases, and
consequently, the concepts of broken gauge symmetry and ODLRO appear,
strictly speaking, not to be applicable in these finite-sized systems.
However, though the phenomena occurred in trapped alkali atoms are not a
thermodynamic phase transition in a rigorous sense, our preceding discussion
might shed some useful light on it. For instance, the relation (\ref{phi})
satisfied by the condensate wave-function, $\Phi _{0}({\bf r})$, which is
normally considered as the order parameter, could also be used to
investigate the linear, not rotational, motion of the condensate to a proper
level of approximation. Quite recently, the irrotational flow and the
quantization of vorticity in a BEC of $^{87}$Rb atoms has been successfully
observed at extremely low temperatures\cite{chevy}.

\acknowledgements
One of authors (GS) would like to acknowledge Profs. H.T. Nieh, C.N. Yang
and B.H. Zhao from whom he learned a great deal about concept of ODLRO. This
work is also supported in part by JSPS, NSFC, State Key Project for Basic
Research of China, and Chinese Academy of Sciences.

\end{document}